\documentclass[12pt,a4paper]{JHEP3}
\usepackage{graphicx}
\def\be{\begin{equation}}
\def\ee{\end{equation}}
\def\bea{\begin{eqnarray}}
\def\eea{\end{eqnarray}}
\def\bei{\begin{itemize}}
\def\eei{\end{itemize}}
\def\bee{\begin{enumerate}}
\def\eee{\end{enumerate}}

\def\sqr#1#2{{\vcenter{\vbox{\hrule height.#2pt
     \hbox{\vrule width.#2pt height#1pt \kern#1pt
           \vrule width.#2pt}
       \hrule height.#2pt}}}}
\def\Box{\mathchoice\sqr55\sqr55\sqr{2.1}3\sqr{1.5}3}

\title{4D gravity on a  brane from bulk higher-curvature terms}

\author{Olindo Corradini 
\\Dipartimento  di Fisica, Universit{\`a} di Bologna 
and  INFN, Sezione di Bologna\\
Via Irnerio, 46 -  
Bologna I-40126, Italy 
\\ E-mail: \email{corradini@bo.infn.it}}

\abstract{We study a gravity model where a tensionful codimension-one 
three-brane is embedded in a bulk with infinite transverse length. 
We find that 4D gravity is induced on the brane already at the classical level
if we include higher-curvature (Gauss-Bonnet) terms in the bulk. Consistency 
conditions appear to require a  negative brane tension as well as a negative 
coupling for the higher-curvature terms.\\[1cm]PACS No.: 04.50+h}

\preprint{hep-th/0405038}

\begin{document}

%
%
\section{Introduction}
In the  context of brane world several mechanisms, that
yield 4D gravity on a brane embedded in  a larger space-time, have been
studied.  Here
we  would like  to present  some  work done on  a codimension-one setup  where
gravity  is  induced  on  the  brane from  bulk  higher-curvature terms.
In particular, we study a gravity model where a bulk five-dimensional
Einstein-Hilbert (EH) action is ameliorated with the Gauss-Bonnet (GB) 
combination of
curvature-squared terms. We also include a bulk cosmological constant 
and a 3-brane $\Sigma$ with generically non-vanishing tension. 
Such a model has been thoroughly investigated over the past few years
under many aspects, often viewed as a 
generalization~\cite{Kim:2000pz,Deruelle:2003tz} 
of the RS setup~\cite{Randall:1999vf}. In particular, 
in~\cite{Deruelle:2003tz} it was shown that, upon adding GB to the 
lowest order action of~\cite{Randall:1999vf}, one obtains the Newtonian 
potential on the (positive tension) brane both in the IR and in the UV 
regime. 
Recently~\cite{Charmousis:2003sq} it was found that 
negative tension solutions in the RS context may present tachyonic 
instabilities, in presence of a bulk GB term. 
Here, we study the aforementioned model under 
a different perspective. First 
of all we allow for solutions that have infinite invariant length in
the transverse direction. In such a situation it is by now well known
that 4D gravity can be induced on the brane through quantum
effects~\cite{Dvali:2000rv,Dvali:2000hr}. At the level of the low-energy 
effective action, if the localized matter is non-conformal, 
loops of matter fields with external gravity lead to a power series 
in derivatives of curvatures, that truncated at the second order amount 
to include on the brane an explicit EH term as well as a contribution  
to the brane tension~\cite{Dvali:2000hr}; it is usually referred to this model
as the Dvali-Gabadadze-Porrati (DGP) model. By now several string theory 
realizations of the DGP model are known~\cite{Kiritsis:2001bc}.

It was also 
pointed out that, upon including a bulk GB term, one can obtain 4D gravity on a
tensionful codimension-2 brane in infinite transverse space, 
even without any explicit EH term on the brane~\cite{Corradini:2001qv}.

In this note we study the occurrence of 4D gravity on a 
tensionful 3-brane in 5D 
bulk via the presence in the bulk of Gauss-Bonnet combination of 
curvatures.~\footnote{In~\cite{Kyae:2003nc} a Gauss-Bonnet correction to the 
tensionless setup of~\cite{Dvali:2000hr} was considered.}  

Hence the model reads~\footnote{For notational convenience we work with a
codimension-one brane on a space-time of unspecified 
dimension $D$, even though we will mostly have in mind  the specific 
case $D=5$.} 
\bea
S = {1\over 2\kappa^2} \int\!\! d^D x\ \sqrt{-G} \Bigl(R 
+\tilde\xi Z - \Lambda \Bigr)
-
\int_\Sigma\!\! d^{D-1} x\ \sqrt{-\hat G} f
\label{eq:model}
\eea 
where
\bea
Z= R^2-4 R_{MN}R^{MN}+R_{MNST}R^{MNST}~,
\label{def:gauss-bonnet}
\eea
is the GB combination and 
\bea
\hat G_{\mu\nu} =\delta_\mu^M \delta_\nu^N G_{MN}\, \Big|_\Sigma ~,
\eea
is the pull-back of the metric $G_{MN}$ on the brane. Although such 
a setup may present interesting cosmological 
features~\cite{Kim:2000pz,Nojiri:2000gv,Dehghani:2004cf}, here we limit our 
study to static solutions.

In order to stress the effect of the presence of the bulk GB term in our setup,
we consider a 
toy-model where the Einstein-Hilbert term is absent, showing that one may get
4D gravity on a codimension-one brane from {\em pure} bulk quadratic terms.

The letter is organized as follows. In section~\ref{section:solutions}
we describe  the type  of backgrounds upon  which gravity  has been
studied.  In   section~\ref{section:planck}  some  preliminary  estimates
concerning   the   parameters   of    the   model   are   given.   In
section~\ref{section:gravity} we study linearized gravity on the static 
solutions described in section~\ref{section:solutions} and then we present 
some comments in the final section.

%
%
\section{Static solutions}
\label{section:solutions}
The equations of motion for the model~(\ref{eq:model}) are given by
\bea
R_{MN}-{1\over 2}G_{MN}\left(R+\tilde\xi Z-\Lambda\right)+2\tilde\xi Z_{MN}
=-{1\over 2} G_{M\rho}G_{N\sigma} \hat G^{\rho\sigma} 
{\sqrt{-\hat G}\over\sqrt{-G}}
\tilde f \delta(z)~,
\eea 
where 
\bea
Z_{MN} = R R_{MN} -2R_{MS} R^S{}_N +R_{MRST}R_N{}^{RST}-2R^{RS}R_{MRNS} 
\eea
and $\tilde f = 2\kappa^2 f$ (in the following we will also identify 
$\tilde\xi = 2\kappa^2 \xi$).

We consider static warped solutions of the kind
\bea
ds^2 = e^{2A(z)}\eta_{MN}dx^M dx^N
\label{eq:ansatz}
\eea
where $z$ parameterizes the transverse direction. The equations of
motion are
\bea
&&
(D-2)(A^{\prime\prime}-{A^\prime}^2)\left[1
-2\tilde\xi(D-3)(D-4){A^\prime}^2e^{-2A}\right]
+{1\over 2} e^{A} \tilde f \delta(z)=0
\label{EoM1}
\\[3mm]
&& (D-1)(D-2)\left[1-\tilde\xi(D-3)(D-4){A^\prime}^2 e^{-2A}\right]{A^\prime}^2
+\Lambda e^{2A} =0
\label{EoM2}
\eea
which, as usual, admit bulk AdS solutions
\bea
A_\pm(z) = - \ln(\pm kz+1)\quad\quad k>0
\eea 
that, in turns, can be combined to give
\bea
A_{1,2}(z) = - \ln(\pm k |z|+1)
\label{solutions}
\eea 
in order to satisfy the jump condition included in~(\ref{EoM1}). 
In fact, the equations of motion become 
\bea
\label{sol:f}
&& \tilde f = \pm 4(D-2)k\left[1-2\tilde\xi(D-3)(D-4)k^2\right]
\\ \label{sol:k2}
&& k^2-\tilde\xi(D-3)(D-4)k^4+{\Lambda\over (D-1)(D-2)} =0
\eea 
and thus
\bea
k^2 = {1\over 2(D-3)(D-4)\tilde\xi
}\left[1+\epsilon\, {\rm sgn}(\xi)\sqrt{1+4\Lambda\tilde\xi 
{(D-3)(D-4)\over (D-1)(D-2)}}\right]\quad\epsilon=\pm 1
\label{k2}
\eea
are the generic bulk solutions for both $A_{1,2}$. The branch characterized by 
$\epsilon=-{\rm sgn}(\xi)$, 
in the limit $\xi\rightarrow 0$ reduces to the pure Einstein solution, and is 
referred to as the EH branch. The other branch ($\epsilon=+{\rm sgn}(\xi)$) is 
called GB branch and is not continuously connected to pure Einstein
solutions. 

The Ricci scalar on the solutions $A_{1,2}$ is given by
\bea
R = -D(D-1)k^2 \pm 4(D-1)k\delta(z)
\label{ricci-0}
\eea
The solution $A_1$ is therefore a GB deformation of the RS2 
solution~\cite{Randall:1999vf} and has finite invariant-length
(compactification volume)
\bea
L\equiv \int^\infty_{-\infty}\!\! dz\ \sqrt{-G} = {2\over (D-1)k}
\eea
although the range of the $z-$coordinate is infinite, as usual. On the 
other hand the lower solution, $A_2$, has infinite invariant-length
\bea
L\equiv \int^{1/k}_{-1/k}\!\! dz\ \sqrt{-G} = +\infty~.
\eea
In general,
the invariant length enters in the expression of the localized Planck mass as
\bea
M_P^{D-3} \sim L M^{D-2} 
\eea 
and thus the solution 2 cannot lead to localization of gravity. 
However the presence of the GB term turns out to play a special role here: 
it will in fact ``induce'' 4D gravity on the brane, along the lines of
what described in~\cite{Corradini:2001qv}. However we will see that the
present codimension-one case is quite peculiar in that positivity
arguments appear to require a negative tension on the brane as well as 
a negative GB coupling.  In the following we 
investigate some features of the aforementioned infinite-length 
solution and  will argue that such apparently odd setup does not
present evident inconsistencies.

%
%
\section{The induced Planck mass and the brane tension}
\label{section:planck}
In order to show that the model~(\ref{eq:model}) reproduces 4D gravity on the 
brane $\Sigma$ also for the infinite-length solution, $A_2(z)$, 
we consider the equations of motion at the linearized level. It is not 
difficult to convince oneself that the GB combination induces a 
(D-1)-dimensional graviton propagator on the brane. The contribution of a term 
quadratic in curvatures to the linearized equations of motion can be 
schematically represented as
\bea
\xi R^{(0)}E_{\mu\nu}^{(1)} \sim \xi k\delta(z)\, \Box_4 h_{\mu\nu}~,
\eea
where $E_{\mu\nu}^{(1)}$ is the linearized Einstein tensor and $R^{(0)}$ is the
distributional part of the zero-th order Ricci scalar given in~(\ref{ricci-0}).
One can thus recognize
\bea
M_P^{D-3}\sim - \xi k
\eea
and therefore the positivity of the induced Planck mass 
requires~\footnote{A model with negative Gauss-Bonnet coupling recently
already appeared in a somewhat different context~\cite{Dehghani:2004cf}.} 
\bea
\xi < 0~.
\eea
 From the jump condition we then see that~\footnote{In the 
tensionless limit equations~(\ref{sol:f}-\ref{sol:k2}) lead to the solitonic 
solution of~\cite{Iglesias:2000iz}. Note that in such a case 
$\xi>0$ and the 
positivity condition on $M_P$ forces to choose the finite-length solution 
$A_1$.}
\bea
f \sim -k[1-2\tilde\xi(D-3)(D-4) k^2] < 0
\eea
and therefore both the tension and the GB coupling turn out to be negative 
in such a set up. Note that this result is valid only for the EH branch; 
in fact, in the GB branch, equation~(\ref{k2}) would lead to negative $k^2$ as 
the coupling $\xi$ is negative. On the other hand, in the EH branch, $k^2 >0 $ 
if the cosmological constant is  negative.

%
%

\section{Gravity on the brane}\label{section:gravity}
In order to have a more precise understanding of the mechanism that induces 
gravity
in this setup let us consider the linearized equations of motion in presence of
 matter localized on the brane.
To do this, let us study small fluctuations around the solution:
\begin{equation}\label{fluctu}
 G_{MN}=\exp(2A)\left[\eta_{MN}+{\widetilde h}_{MN}\right]~,
\end{equation}
where for convenience reasons we have chosen to work with 
${\widetilde h}_{MN}$
instead of metric fluctuations $h_{MN}=\exp(2A){\widetilde h}_{MN}$. We will 
use the following notation
\begin{equation}
 H_{\mu\nu}\equiv{\widetilde h}_{\mu\nu}~,~~~A_\mu\equiv
 {\widetilde h}_{\mu D}~,~~~\rho\equiv{\widetilde h}_{DD}~.
\end{equation}
for the component fields. The coupling between the localized matter and the 
graviton
field reads
\begin{equation}\label{int}
 S_{\rm int}={1\over 2} \int_\Sigma d^{D-1} x ~T_{\mu\nu}
 H^{\mu\nu}~.
\end{equation} 
We thus get the following set of equations
\begin{eqnarray}\label{EOM1oGB}
 &&\left[1-2(D-3)(D-4)\tilde\xi {A^\prime}^2 e^{-2A}\right]\Biggl\{
 \partial_\sigma\partial^\sigma 
 H_{\mu\nu} +\partial_\mu\partial_\nu H
 -2\partial_{(\mu} \partial^\sigma H_{\nu)\sigma}-\nonumber\\
 &&\eta_{\mu\nu}
 \left(\partial_\sigma\partial^\sigma H-\partial^\sigma\partial^\rho
 H_{\sigma\rho}\right)+
 H_{\mu\nu}^{\prime\prime}-\eta_{\mu\nu}H^{\prime\prime}+
 (D-2)A^\prime\left(H_{\mu\nu}^\prime-\eta_{\mu\nu}H^\prime\right)-
 \nonumber\\
 &&2\left[\partial_{(\mu} A_{\nu)}^\prime  -
 \eta_{\mu\nu}\partial^\sigma A_\sigma^\prime+ (D-2)A^\prime
 \left(\partial_{(\mu} A_{\nu)} 
 -\eta_{\mu\nu}\partial^\sigma
 A_\sigma\right)\right]\nonumber+\\
 &&\left[\partial_\mu\partial_\nu\rho-\eta_{\mu\nu}
 \partial_\sigma\partial^\sigma 
 \rho+\eta_{\mu\nu}\left((D-2)A^\prime\rho^\prime
 +(D-1)(D-2){A^\prime}^2\rho\right)\right]\Biggr\}-\nonumber\\
 &&4(D-4)\tilde\xi\left(A^{\prime\prime}-{A^\prime}^2\right) e^{-2A}
 \Biggl\{\partial_\sigma\partial^\sigma 
 H_{\mu\nu} +\partial_\mu\partial_\nu H
 -2\partial_{(\mu} \partial^\sigma H_{\nu)\sigma}-\nonumber\\
 &&\eta_{\mu\nu}
 \left(\partial_\sigma\partial^\sigma H-\partial^\sigma\partial^\rho
 H_{\sigma\rho}\right)+
 (D-3)A^\prime\left[H_{\mu\nu}^\prime-2\partial_{(\mu} A_{\nu)}
 -\eta_{\mu\nu}\left(H^\prime-2\partial^\sigma A_\sigma
 \right)\right] \Biggr\}+
\nonumber\\
&&2(D-2)\left(A^{\prime\prime}-{A^\prime}^2\right)
\left[1-4(D-3)(D-4)\tilde\xi {A^\prime}^2 e^{-2A}\right]\eta_{\mu\nu}\rho
\nonumber\\
&&=-2\kappa^2 \left( T_{\mu\nu}+{1\over 2}\eta_{\mu\nu}f\rho\right) 
\delta(z)~,\\[5mm]
 \label{EOM2oGB} 
 &&\left[1-2(D-3)(D-4)\tilde\xi {A^\prime}^2 e^{-2A}\right]\Biggl[
 \left(\partial^\mu H_{\mu\nu}-\partial_\nu H\right)^\prime 
 -\partial^\mu F_{\mu\nu}+\nonumber\\
 && (D-2)A^\prime \partial_\nu\rho\Biggr]
=0~,\\[5mm]
\label{EOM3oGB}
 &&\left[1-2(D-3)(D-4)\tilde\xi {A^\prime}^2 e^{-2A}\right]
 \Biggl[-\left(\partial^\mu\partial^\nu H_{\mu\nu}-\partial^\mu\partial_\mu H
 \right) +\nonumber\\
 &&(D-2) A^\prime \left(H^\prime-2\partial^\sigma A_\sigma\right)
 -(D-1)(D-2) {A^\prime}^2\rho\Biggr]
=0~.
\end{eqnarray}
The {\em graviphoton} $A_\mu$ can be set to zero everywhere. The reason is 
twofold; on
the one hand $A_\mu(z)$ is $Z_2$-odd and therefore it vanishes at $z=0$; then 
using 
the unbroken diffeomorphisms $A_\mu$ can be gauged away 
completely~\cite{Kakushadze:2000ix}. On the other hand $A_\mu$ can couple to 
brane matter only through $\partial_\mu T^{\mu\nu}$ which vanishes for 
conserved brane matter.
  
Hence, setting such field to zero, eq.~(\ref{EOM2oGB}) reduces to
\bea 
\label{EOM2oGB+} 
 &&\left(\partial^\mu H_{\mu\nu}-\partial_\nu H\right)^\prime 
 + (D-2)A^\prime \partial_\nu\rho =
 0
\eea
and can be solved to give
\bea
\left(H_{\mu\nu}-\eta_{\mu\nu}H\right)^\prime+(D-2)A^\prime\eta_{\mu\nu}\, \rho
=\eta_{\mu\nu} F(z)~.
\eea
It is not difficult to see that $F\equiv 0$ as the fluctuations are 
expected to vanish asymptotically. Therefore, using the background equation
of motion~(\ref{EoM1}), equation~(\ref{EOM1oGB}) can be cast in the form
 \begin{eqnarray}\label{EOM1oGB+}
 &&\left[1-2(D-3)(D-4)\tilde\xi {A^\prime}^2 e^{-2A}\right]\Biggl\{
 \partial_\sigma\partial^\sigma 
 H_{\mu\nu} +\partial_\mu\partial_\nu H
 -2\partial_{(\mu} \partial^\sigma H_{\nu)\sigma}-\nonumber\\
 &&\eta_{\mu\nu}
 \left(\partial_\sigma\partial^\sigma H-\partial^\sigma\partial^\rho
 H_{\sigma\rho}\right)+
 H_{\mu\nu}^{\prime\prime}-\eta_{\mu\nu}H^{\prime\prime}-
 \left[\partial_\mu\partial_\nu\rho-\eta_{\mu\nu}
 \partial_\sigma\partial^\sigma 
 \rho+\right.\nonumber\\
&&\left.\eta_{\mu\nu}(D-2)\left(A^\prime\rho^\prime
 +{A^\prime}^2\rho\right)\right]\Biggr\}-
 4(D-4)\tilde\xi\left(A^{\prime\prime}-{A^\prime}^2\right) e^{-2A}
 \Biggl\{\partial_\sigma\partial^\sigma
 H_{\mu\nu}+
\nonumber\\
&& \partial_\mu\partial_\nu H
 -2\partial_{(\mu} \partial^\sigma H_{\nu)\sigma}-
\eta_{\mu\nu}
 \left(\partial_\sigma\partial^\sigma H-\partial^\sigma\partial^\rho
 H_{\sigma\rho}\right) \Biggr\}
=-2\kappa^2 \hat T_{\mu\nu} \delta(z)~,\\[5mm]
\label{EOM3oGB+}
 && -\left(\partial^\mu\partial^\nu H_{\mu\nu}-\partial^\mu\partial_\mu H
 \right) +
 (D-2) A^\prime H^\prime
 -(D-1)(D-2) {A^\prime}^2\rho =0~,
\end{eqnarray} 
where
\bea
\hat T_{\mu\nu} = T_{\mu\nu} -{1\over 2}\eta_{\mu\nu}f\rho~.
\eea
In~(\ref{EOM1oGB+}) the terms in the curly bracket multiplying
\bea
\left(A^{\prime\prime}-{A^\prime}^2\right)e^{-2A}=2k \delta(z)
\eea
are nothing but the linearized Einstein tensor in (D-1)-dimensional 
flat space. Therefore it is easy to recognize 
\bea
M_P^{D-3} = - 8(D-4)k\xi
\label{GB-Planck}
\eea
as the induced Planck mass on the brane. The leftover contribution 
in~(\ref{EOM1oGB+}) is the bulk graviton propagator with an effective mass
\bea
2\kappa^2 M_{\rm eff}^{D-2} &=&  
1-2(D-3)(D-4)\tilde\xi {A^\prime}^2 e^{-2A} 
=1-2(D-3)(D-4)\tilde\xi k^2 ~.
\eea
In other words, the linearized equations of 
motion~(\ref{EOM1oGB+}-\ref{EOM3oGB+})  for the spin-two 
fluctuations, are consistent with those associated to the model
\bea
S_* = M_{\rm eff}^{D-2} \int\!\! d^Dx \sqrt{-G}R 
+M_P^{D-3}\int_\Sigma\!\! d^{D-1} x \sqrt{-\hat G}\hat R
+{1\over 2} \int_\Sigma\!\! d^{D-1} x T_{\mu\nu} H^{\mu\nu}~. 
\label{eq:model-eq}
\eea
At the linearized level the 
model~(\ref{eq:model-eq}) coincides with the model~(\ref{eq:model}),  
albeit they differ beyond the linearized approximation.
It is worth noting that, at the linearized level, 
the only effect of the GB combination in the bulk is
the ``renormalization'' of the Planck mass. If fact, let us stress
that one would 
obtain a canonical bulk propagator with positive 
nonvanishing bulk Planck mass even {\em without} 
the bulk Einstein-Hilbert term, that is starting from the purely 
higher-derivative model
\bea
S_{GB} = \int\!\! d^D x\ \sqrt{-G} \Bigl(
\xi Z - V_B \Bigr)
-\int_\Sigma\!\! d^{D-1} x\ \sqrt{-\hat G} f~.
\label{eq:model-GB}
\eea 
Such a Planck mass would be
\bea
M_{GB}^{D-2} = -2(D-3)(D-4) \xi k^2 
\eea
that is positive as the coupling $\xi$ is negative in this model. In
this limit, where the bulk EH term is absent
\bea
k = \left[ V_B\over (D-1)\cdots (D-4) \xi\right]^{1\over 4}
\label{k-GB}
\eea
is the expression of the AdS scale as can be easily inferred 
from~(\ref{sol:k2}), and the brane Planck mass is again given 
by~(\ref{GB-Planck}). This limit is unphysical as 
it is of course not obvious that higher curvature contributions would not
spoil the results. However, we presented it here as we think it might be
helpful to understand better how 4D gravity is induced on the brane in the
setup  we have considered.

Before ending  this section  we would like  to point out  some crucial
differences  with  the  results  obtained  recently  by  Deruelle  and
Sasaki~\cite{Deruelle:2003tz}.         The  setup studied
in~\cite{Deruelle:2003tz} differs from the one studied here in that 
they  considered a RS2 type  of background, that  is, 4D flat
space  with a noncompact  warped finite-length  extra dimension  and a
positive-tension  3-brane.   In such  a  case  two  mechanisms are  at
work. On the one hand, there is a Randall-Sundrum type of localization
that yields 4D  gravity on the brane at large  scales ($r >> r_1\equiv
1/k$).   On  the  other  hand   there  is  a  Brane  Induced  Gravity 
(BIG)
mechanism, given there by the bulk GB, that yields 4D gravity at small
scales ($r  << r_2 \sim \xi  k/M^{D-2}$). Hence, in  the BIG dominant
regime  the crossover  scale $r_2$  covers the  other  crossover scale
$r_1$, and therefore  perturbative gravity is 4D at  {\em all} scales,
as  pointed  out  in~\cite{Kiritsis:2002ca}.  In  our model  the
invariant  length  of  the   transverse  direction  is  infinite,  and
therefore  we have  no  RS localization.  We  thus have  a single  BIG
crossover scale
\bea  
r_C \equiv {-8(D-4)\xi  k\over M^{D-2}-2(D-3)(D-4)\xi k^2}
\eea 
and  gravity is four  dimensional at distances much  smaller than
the  previous scale.

%
%
\section{Comments}
\label{section:comments}
The aim of this letter was to show that gravity can be ``induced'' on a
3-brane embedded in a 5D bulk with infinite transverse invariant length, 
via the presence of a bulk GB combination. In fact, for the only purpose of
further underlining this mechanism, we have shown 
that gravity on the brane remains 4D in the ultraviolet regime even in a
toy-model with purely GB bulk gravity. 
In~\cite{Charmousis:2003sq} it was shown that negative-tension branes in the
RS+GB context project-in a normalizable tachyon and project-out a 
(normalizable) zero-mode, 
thus leading to an instability of the setup. We claim that such
 arguments of instability do not apply
in our case as gravity here is not localized or, in other words, there is no 
(normalizable) 4D graviton zero mode~\cite{Dvali:2000rv}.

As to the negativity of the GB coupling, note first that $\xi <0 $ yields a 
positive contribution to the bulk kinetic term, so that there is no fear of
bulk ghost fields. On the other hand, from string theory there is 
{\em a priori} no stringent constraint on the sign of the GB coupling. In fact,
 if on the one hand the
sign of the Gauss-Bonnet combination in 10-dimensional heterotic string theory
is positive~\cite{Gross:1986mw}, on the other hand in compactified 
theories such sign might depend on the details of the compactification. In 
fact, the Gauss-Bonnet combination in  5D can be, for instance, obtained 
from the compactification of M-theory on $CY_3$~\cite{Antoniadis:1997eg}. In 
such a case the coupling constant for the GB term is moduli dependent and its 
sign is {\em a priori} not fixed. 

Finally, a comment is in order regarding models with generic (non GB) 
combinations of quadratic curvature terms. Such models do not appear to admit
solutions with delta-function type of discontinuities (thin sources), like the 
ones presented here. On the other hand it is plausible to conjecture that they 
might induce gravity on smooth types of branes. It is also plausible that such
mechanism might persist in higher (larger than two) codimension setups. 
Curvature-squared terms were shown to be helpful to smooth out 
singularities in  higher-codimension brane worlds with infinitely large 
transverse space~\cite{Corradini:2002es}; they could also be beneficial to 
induce gravity in such setups. A thorough investigation in this direction is 
still missing.

%
%
\acknowledgments{The work has been partly supported by the EC commission via
  the FP5 grant HPRN-CT-2002-00325 and by a Marco Polo fellowship of the
  University of Bologna. The author  is grateful to 
G.L.~Alberghi, F.~Bastianelli, C.~Charmousis, A.~Fabbri, S.~Davis,
  E.~Kohlprath, A.~Iglesias, P.~Langfelder, P.~Vanhove, C.~Zoubos and 
especially E.~Kiritsis 
for help and discussions. He would also like to
thank the CPHT of the \'Ecole Polytechnique for their 
kind hospitality while parts of this work were completed.}


%
%

\end{document}